\documentclass[
 reprint,
 superscriptaddress,
showpacs,preprintnumbers,
 amsmath,amssymb,
 aps,
 prl,
 longbibliography
]{revtex4-1}

\usepackage{mathtools}
\usepackage{graphicx}
\usepackage{subfigure}
\usepackage{multirow}
\renewcommand{\thesubfigure}{\alph{subfigure}}
\makeatletter
\renewcommand{\@thesubfigure}{(\thesubfigure)\space}
\makeatother
\usepackage{dcolumn}
\usepackage{bm}
\usepackage[colorlinks, linkcolor=blue, citecolor=blue, urlcolor=blue]{hyperref}

\begin{document}


\title{Critical Scaling Behaviors of Entanglement Spectra}

\author{Qicheng Tang}

\author{W. Zhu}

\affiliation{%
	School of Science, Westlake University, Hangzhou 310024, China and\\
	Institute of Natural Sciences, Westlake Institute of Advanced Study, Hangzhou 310024, China
}%

\date{\today}

\begin{abstract}
	We investigate the evolution of entanglement spectra under a global quantum quench from a short-range correlated state to 
	the quantum critical point. Motivated by the conformal mapping, we find that the dynamical entanglement spectra
	demonstrates distinct finite-size scaling behaviors from the static case.
	As a prototypical example, we compute real-time dynamics of the entanglement spectra of a one-dimensional transverse-field Ising chain. 
	Numerical simulation confirms that, 
	the entanglement spectra scales with the subsystem size $l$ as $\sim l^{-1}$ for the dynamical equilibrium state, 
	much faster than  $\propto \log^{-1} l$ for the critical ground state.
	In particular, as a byproduct, the entanglement spectra at the long time limit faithfully gives universal tower structure of underlying Ising criticality,
	which shows the emergence of operator-state correspondence in the quantum dynamics.
\end{abstract}

\maketitle


Conformal field theory (CFT)~\cite{DiFrancesco639405} has become a profitable tool as a 
diagnosis of critical phenomena in two dimensional statistical models.
In the equilibrium case, the conformal invariance at the critical point sets rigid constrains on
physical properties by a set of conformal data including the central charge, conformal dimensions and operator product expansion coefficients.
In the past decades, great success has been achieved in condensed matter physics, 
especially for minimal models with a finite number of primary scaling operators (irreducible representations of the Virasoro algebra)
~\cite{belavin1984infinite1, belavin1984infinite2, friedan1984conformal, CARDY1986186, CARDY1986200}.

In general, due to gaplessness nature massive entanglement should play a vital role at or around the critical point.
One remarkable achievement is \cite{srednicki1993entropy, HOLZHEY1994443, Calabrese_2004, korepin2004universality, Calabrese_2005, calabrese2006time, fradkin2006entanglement, 
Calabrese_2007, calabrese2008entanglement, hsu2009universal, Calabrese_2009, Nienhuis_2009, alba2010entanglement, calabrese2010parity, calabrese2012entanglement, calabrese2013entanglement, 
cardy2014thermalization, calabrese2014finite, Coser_2014, Cardy_2016review, Calabrese_2016, Cardy_2016, Alba_2017, Wen_2018, giudici2018entanglement, di2019entanglement,vidal2003entanglement,pollmann2009theory}, 
CFT provides a novel way to connect the quantum entanglement and critical phenomena. 
It is found that the conformal invariance in critical ground states results in a universal scaling of the entanglement entropy depending on the central charge $c$ 
~\cite{HOLZHEY1994443,  korepin2004universality, Calabrese_2004, fradkin2006entanglement, hsu2009universal}.
Interestingly, by extending this idea, the entropy can be applied to identify quantum critical points in higher dimensions 
~\cite{fradkin2006entanglement, hsu2009universal, MAX2009, Seth2017, WZhu2018}. 
Besides the entropy, other entanglement-based measures also attract lots of attention. 
The eigenvalues of reduced density matrix, called entanglement spectrum (ES), is such an example,
which contains much richer information than the entropy ~\cite{li2008entanglement, LAFLORENCIE20161}.
In addition to the evidences in topological gapped systems~\cite{fidkowski2010entanglment, prodan2010entanglement, turner2010entanglement, qi2012general}, 
the ES is also proposed to describe the quantum critical point 
~\cite{calabrese2008entanglement, thomale2010nonlocal, de2012entanglement, lepori2013scaling, giampaolo2013universal, lundgren2016universal, schuler2016universal, whitsitt2017spectrum, Vid2015}.
However, compared to the well-established boundary law for gapped states,
much less is known about the critical behavior of the ES ~\cite{lauchli2013operator, Laflorencie_2014},
which casts doubt on direct application of the ES in the critical systems.

Beyond equilibrium, quantum dynamics attracts considerable attention recently, particular in approaching to steadiness and thermalization.
Universal entanglement structures are expected to leave some marks in the dynamic process, e.g., 
central charge $c$ controls the growth of entropy
\cite{Calabrese_2005, Calabrese_2009}.
However, novel example \cite{HOLZHEY1994443} is still rare, and 
to extract the conformal data in microscopic models is a challenging task~\cite{Calabrese_2007, Coser_2014, Calabrese_2016, di2019entanglement}.

In this paper we present a systematical study of dynamics of the ES in the process of quantum quench.
Inspired by the CFT, we compute the real-time dynamics of 1D transverse-field Ising (TFI) model,
through a protocol by quenching from a gapped state to the critical point. 
We successfully establish that, quantum quench dynamics indeed encodes universal signatures of quantum critical point. 
First, the ES at long time dynamics converges to the CFT operators as $\propto l^{-1}$, 
which is much faster than that for critical ground state as $\propto \log^{-1} l$.
Second, fast convergence allows us to extract conformal information including 
conformal dimensions and related degeneracy, which 
unambiguously pin down the underlying nature of quantum critical point ($c=1/2$ Ising theory in our case).
These key findings open a door to extract quantum criticality in many-body dynamics.

\begin{figure}
	\includegraphics[width=\columnwidth]{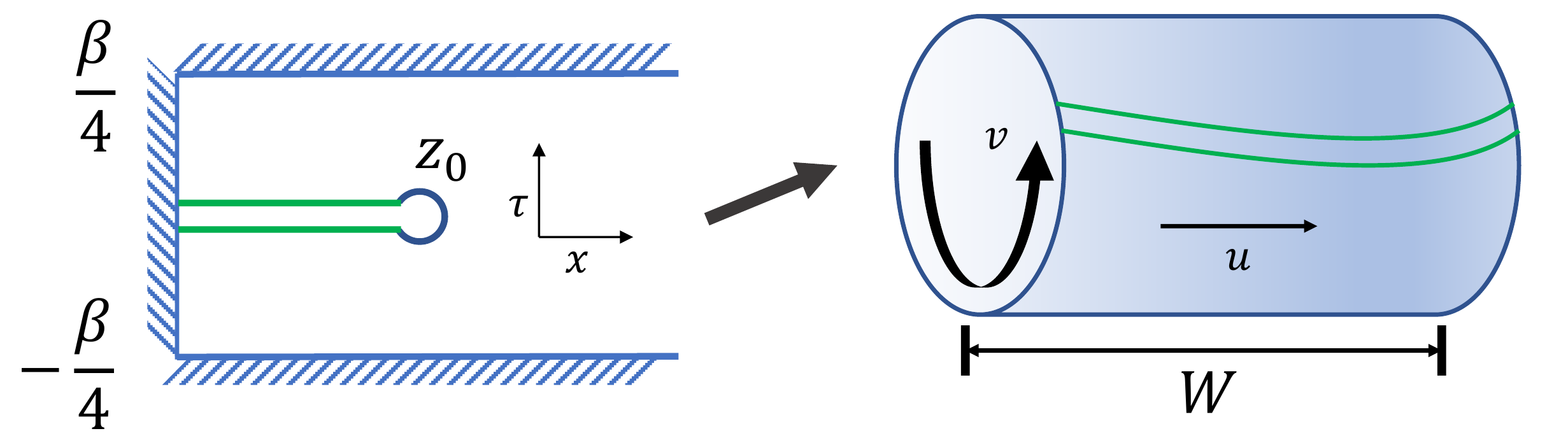}
	\caption{\label{fig:geometry} Conformal mapping from the semi-infinite strip in $z$-plane to the annulus in $w$-plane. 
		(left panel) The global quench of a semi-infinite short-range correlated chain (with correlation length $\beta$) is considered as a boundary CFT with 
		$\tau \in [-\frac{\beta}{4}, \frac{\beta}{4}]$. The green lines represent the branch cut $C=\left\{z=x+i\tau; x \in [0,L], \tau \in [-\frac{\beta}{4}, \frac{\beta}{4}] \right\}$. 
		(right panel) The annulus after conformal mapping $w=f(z)$, where the branch cut $C$ is mapped to a $f(C)$ which connects the two edges of the annulus (see Ref. ~\cite{sm} for details). 
		The circumference along $v={\rm Im}w$ direction is $2\pi$, and the width of the annulus along the $u={\rm Re}w$ direction is denoted by $W$.} 
\end{figure}

\textit{Entanglement spectrum in two dimensional conformal field theory.---}\label{sec:CFT}
We consider the global quantum quench to a critical point that is governed by a CFT ($H(t>0)=H_{\rm CFT}$), and the initial state $|\psi_0\rangle$ is chosen to be the ground state of a gapped Hamiltonian $H(t=0)=H_0$.
In this work, we study the geometry of a semi-infinite chain.
In boundary CFT ~\cite{Cardy_2016, Calabrese_2016, Wen_2018}, the corresponding time-dependent density matrix $\rho(t)=e^{-iHt}|\psi_0\rangle \langle\psi_0|e^{+iHt}$ 
can be represented geometrically by a semi-infinite strip in complex plane with width $\tau={\rm Im}(z) \in [-\frac{\beta}{4}, \frac{\beta}{4}]$ and 
length $x={\rm Re}(z) \in [0, +\infty)$.
The entanglement dynamics of a finite subsystem $A\equiv [0,l]$ is our focus.
The reduced density matrix $\rho_A(t)$ can be calculated by sewing together the points which are not in $A$ (the geometric distribution of partial trace).
This can be achieved by conformally mapping the semi-infinite strip in $z$-plane onto the annulus in $w$-plane, with $w=f(z)$, as shown in Fig.~\ref{fig:geometry}.
On the annulus, the entanglement Hamiltonian $H_E = -\log \rho_A$ can be considered as the generator of translation along the $v={\rm Im}w$ direction, 
i.e. $H_E$ has the same structure with the CFT Hamiltonian $H_{\rm CFT}$.
Then $H_E$ in original $z$-plane can be calculated by mapping the CFT Hamiltonian back to the $z$-plane~\cite{Cardy_2016,Wen_2018}.
In this context, one can obtain the entanglement Hamiltonian for a finite interval $A$ of $[0, l]$  in a semi-infinite critical ground state as (in the static case)
\begin{equation}\label{eq:GSHE}
H_E (t=0)= \int_A \frac{\pi(l^2-x^2)}{l} H_{\rm CFT}(x)dx \ .
\end{equation}
While in the dynamic case, we figure out the entanglement Hamiltonian in the long-time limit in the process of quantum quench as \cite{Cardy_2016,Wen_2018} (details see Ref. \cite{sm})
\begin{equation}\label{eq:DSSHE}
H_E(t \to \infty) \simeq {\beta}\int_A H_{\rm CFT}(x) dx \ .
\end{equation}
Here we stress that,  
although $H_E$ depends on the CFT Hamiltonian density $H_{CFT}(x)$ in both static and dynamic case, 
the entanglement Hamiltonian has an additional spatial-dependent envelope function $l^2-x^2$ in the static case (Eq. \ref{eq:GSHE}), 
which is in sharp contrast to that in dynamic case (Eq. \ref{eq:DSSHE}).
This difference will lead to the influence on the static ES, as we will discuss below.

Another notable difference is, the ES has distinct dependence on
the subsystem size $l$.
In CFT, the width of the annulus $W$ along the $u={\rm Re}w$ direction plays important role in the scaling behavior of ES, 
through
\begin{equation}\label{eq:ES_W}
E_i - E_j = \frac{2\pi^2(\Delta_i - \Delta_j)}{W}
\end{equation}
where $E_i$ is the ES that is the eigenvalues of $H_E$, and $\Delta_i$ is conformal dimension of the CFT.
The width $W$ can be obtained by a straightforward calculation $W = {\rm Re} \mathcal{W} = \frac{1}{2}(\mathcal{W} + \overline{\mathcal{W}})$ with $\mathcal{W} = f(i\tau+L-\epsilon) -f(i\tau)$.
For critical ground states on a semi-infinite chain, one obtains the following dependence in the static case
\begin{equation}
W(t=0) = 2\log\frac{l}{\epsilon},\label{eq:Wstatic}
\end{equation}
$\epsilon$ is a scale relevant cut-off.
This makes the entropy at the critical point $S \propto \log l$, and the ES proportional to $E_i-E_j \propto \log^{-1} l$.
In particular, in the case of global quenching a semi-infinite chain considered in present work,
the width $W$ shows the following behaviors \cite{Cardy_2016,Wen_2018,sm}:
\begin{equation}\label{eq:Wdynamic}
W(t>0) \sim
\begin{cases}
\frac{2\pi}{\beta}t \ , \qquad  t < l \\
\frac{2\pi}{\beta}l \ , \qquad  t \to \infty
\end{cases}
\end{equation}
Hence, the ES of dynamical equilibrium state approximates to the CFT scaling spectrum with speed $\propto l^{-1}$.
Moreover, one can obtain that $W$ approaches steadiness exponentially after the saturated time $t=l$ as \cite{sm}
\begin{equation}\label{eq:expWt}
W(t>l) \simeq  W(t \to \infty) - \frac{1}{2} \exp [-\frac{4\pi(t-l)}{\beta}] \ ,
\end{equation}
and it is also reflected in the dynamics of EE and ES.

\textit{Numerical Results.---}
We test the CFT prediction of entanglement dynamics in TFI chain with open boundary condition
\begin{equation}
H_{\rm TFI} = -\sum_{i} \sigma^x_i \sigma^x_{i+1} - g \sum_{i} \sigma^z_i
\end{equation}
where $\sigma^x_i, \sigma^z_i$ are Pauli matrices at $i$-th site, and $g$ is the strength of the magnetic field.
There exists a quantum phase transition between the ferromagnetic ($g<1$) and paramagnetic ($g>1$) phases, and the ground state is gapless only at the critical point $g=1$.
The critical ground state of TFI chain is described by the minimal model with central charge $c=\frac{1}{2}$, and the corresponding scaling operators are shown in Tab. \ref{tab:op}.
TFI chain can be solved exactly by introducing Jordan-Wigner transformation~\cite{LIEB1961407}, and its ES can be calculated from the \emph{correlation matrix}~\cite{Peschel_2009}.
We consider the ES dynamics during global quench from a ground state of gapped phase of TFI chain ($g \neq 1$) to the Ising CFT ($g=1$).
The time-dependent ES can be  calculated by the time-dependent correlation matrix ~\cite{Torlai_2014}.
In this calculation, the total system size is up to $L=1024$, and we choose subsystem size $l \ll L$. 
In the condition of $l \ll L$, when we consider the time domain $t<L$ we can safely neglect the boundary effect from $x\sim L$ on the subsystem $A=[0,l]$.

In addition, we also solve TFI model using the matrix product state approach,
i.e. the time-evolving block decimation (TEBD) technology \cite{Vidal2004}.
The bond dimension is adopted by 1024, and the truncation error is set to be $10^{-8}$.
For dynamics, the time evolution operator is approximated by using second order Trotter-Suzuki decomposition, and the time step is chosen to be $dt=0.01$.
Under the current convergence criterion, the total system sizes in TEBD calculations are limited to $L\le 72$.

\begin{figure}
	\centering
	\includegraphics[width=0.8\columnwidth]{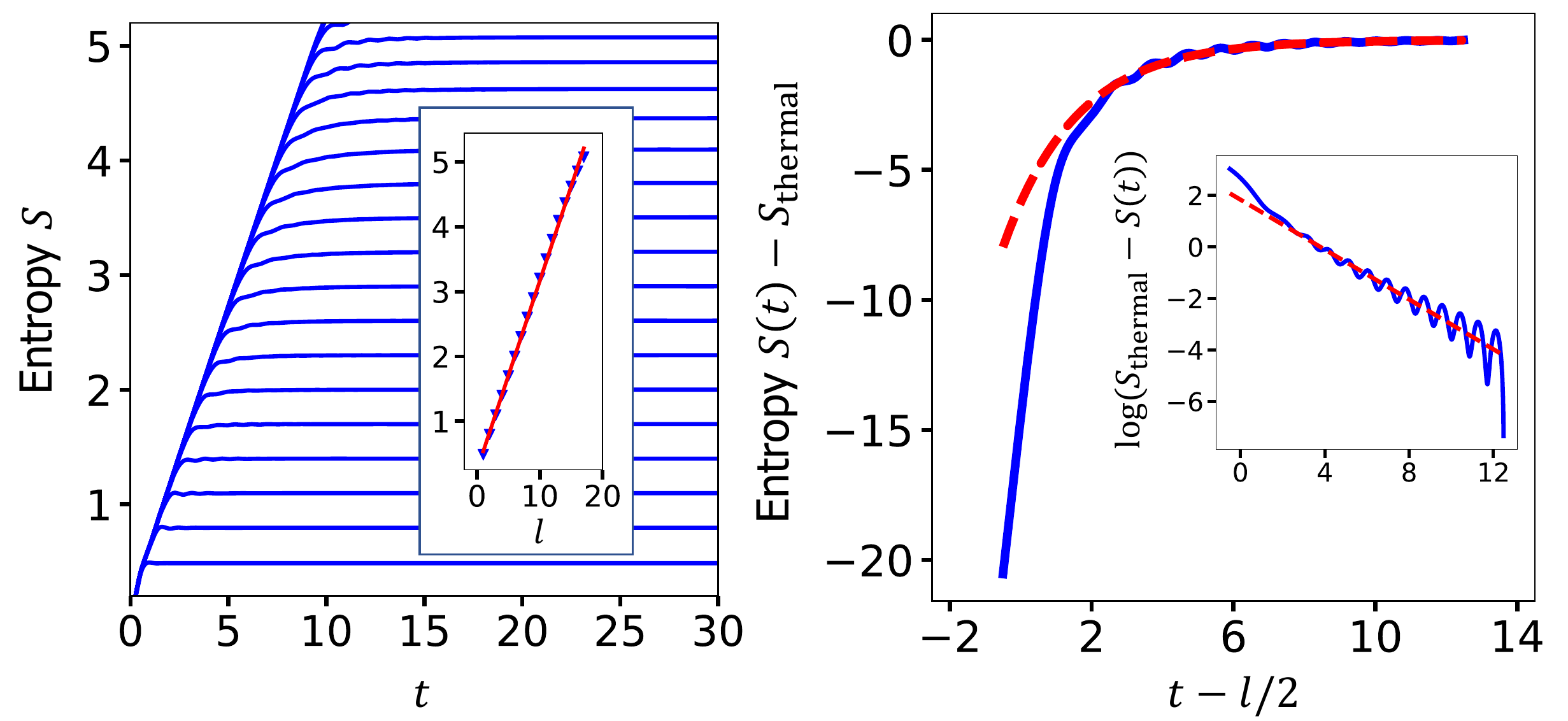}
	\caption{ Entanglement entropy evolution during global quench from gapped phase ($g=4$) to critical point ($g=1$) with total system size $L=512$.
	(left panel) Entropy $S(t)$ grows linearly with the ``entanglement velocity'' $v_E \approx 0.52$. (inset) The finite-size saturation leads to a volume law entropy in dynamic equilibrium state.
	A linear fit (red line) of $S(l)=al+b$ for the numerical data (blue triangle) gives $a=0.29$ and $b=0.24$.
	(right panel) Entropy $S(t)-S_{\rm thermal}$ for $l=15 \ll L=512$ exhibits exponential dependence on $t-l/2$ after reaching the typical time $t \sim l/2$.
	(Inset) A linear fit (red dash lines) of $\log[S_{\rm thermal}-S(t)]=a'(t-l/2)+b'$ for the numerical data (blue lines) gives $a'=-0.4844 \pm 0.0078$ and $b'=1.833 \pm 0.060$.}\label{fig:EE}
\end{figure}

In Fig.~\ref{fig:EE}, we present numerical result of the EE evolution during global quench in TFI chain with total system size $L=512$.
As shown in the left panel, the EE grows linearly with the ``entanglement velocity'' at early times, 
then saturated by finite subsystem size $l$ at time $t = l/2$.
The inset provides numerical evidence of the resulting volume law by a linear fitting of EE with $l$.
The right panel of Fig.~\ref{fig:EE} shows an exponential approaching to equilibrium at late times, 
consistent with Eq. \ref{eq:expWt}.

Now we turn to consider the ES, and
we will address how the ES represents the scaling operators in CFT.  
In CFT, the scaling operators (representations of the corresponding Virasoro algebra) can be obtained by \emph{$q$-expansion} of the partition function.
For Ising model with open (free) boundary condition, one can find that there are only two primary scaling operators: 
the identity $I$ with conformal dimension $\Delta=0$ and the energy density $\epsilon$ with $\Delta=\frac{1}{2}$~\cite{CARDY1986186, CARDY1986200}.
Their characters can be expansed as following
\begin{equation}
\begin{aligned}
\chi_I \ = \ & q^{-1/48}(1+q^2+q^3+2q^4+2q^5 \\
& \qquad \quad +3q^6+3q^7+5q^8+\cdots) \\
\chi_{\epsilon} \ = \ & q^{1/2-1/48}(1+q+q^2+q^3+2q^4+2q^5 \\
& \qquad \qquad +3q^6+3q^7+5q^8+\cdots), \\
\end{aligned}
\end{equation}
which gives the spectrum of the primary operators and corresponding descendants, each expansion term $mq^n$ represents $m$-fold degenerate scaling operators with conformal dimension $\Delta_i=n$.
The \emph{operator-state correspondence} suggests that the eigenspectrum of (entanglement) Hamiltonian shares the structure of scaling operators, 
and this relation has been well studied in the energy spectrum of spin chains.~\cite{chepiga2017excitation, milsted2017extraction, zou2018conformal, zou2019emergence}.

\begin{figure}
	\centering
	\includegraphics[height=5.5cm]{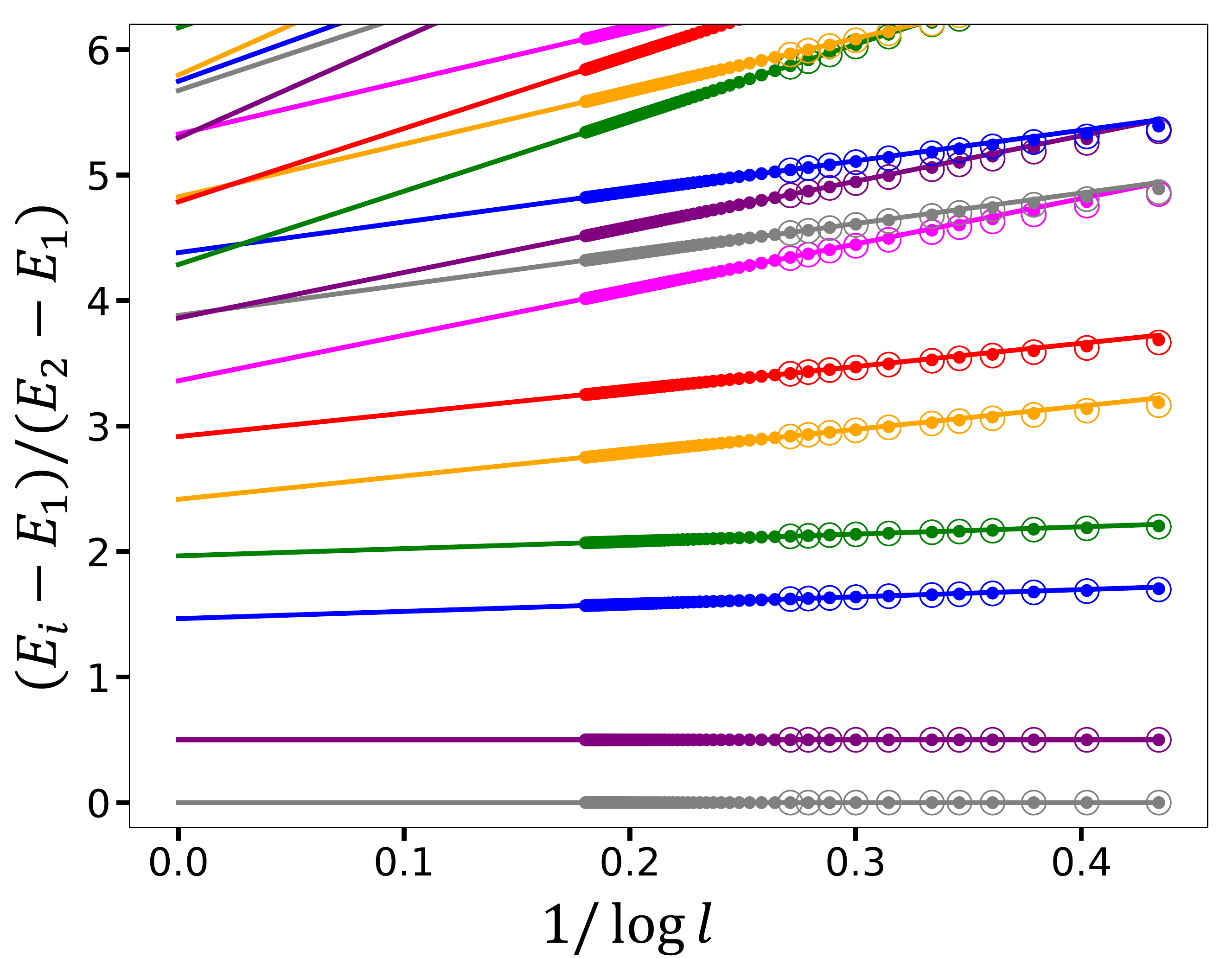} 
	\includegraphics[height=5.5cm]{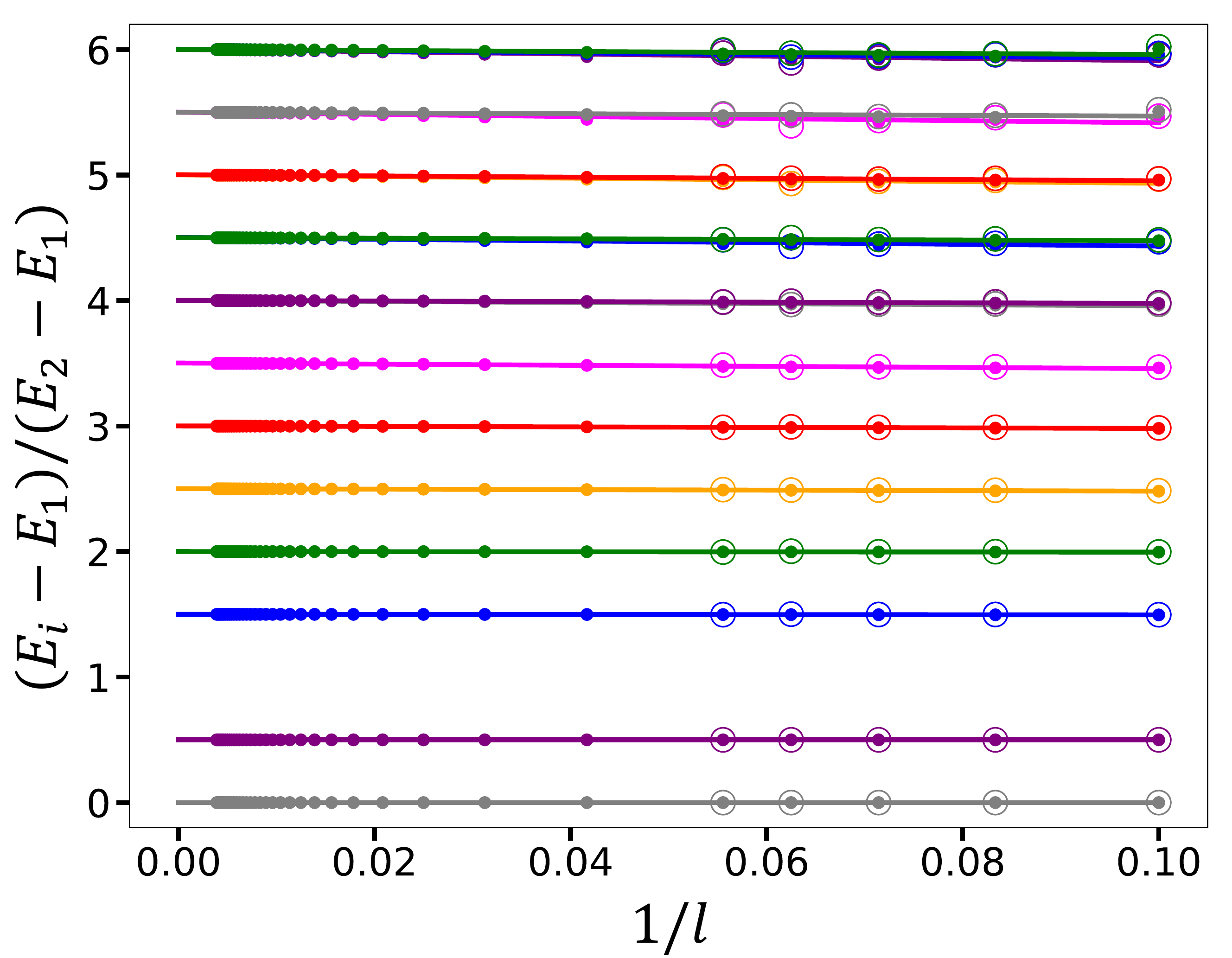}
	\caption{\label{fig:scaling_ES} Entanglement spectra of TFI model as a function of the subsystem size $l$. Here we set the total system size $L=4l$. 
		The solid dots and open circles label the exact solutions and TEBD results, respectively.
		(top) Entanglement spectra for critical ground state ($g=1$), and the related fitting lines in the form of $a\log^{-1} l+b$. 
		(bottom) Entanglement spectra for global quench from the gapped phase ($g=4$) to critical point ($g=1$). 
		The best fitting lines are in form of $al^{-1}+b$. 
		} 
\end{figure}

The numerical results of the ES of TFI chain for the critical ground state and dynamical equilibrium state are respectively plotted in Fig.~\ref{fig:scaling_ES},
which includes the main result of the current work.
Here we consider the entanglement cut always located at $l = L/4$ with changing the total system size $L$.
In order to make a direct comparison to the Ising scaling spectrum, 
we renormalize ES by setting the lowest level to $0$ and the second level to $1/2$, 
and show the quantities $(E_i-E_1)/(E_2-E_1)$ in Fig. \ref{fig:scaling_ES}. Several salient features are found in the ES.
First, we identify distinct scaling behaviors of the ES depending on subsystem size $l$, 
by comparison the static ground state and dynamic equilibrium state. 
As shown in Fig. \ref{fig:scaling_ES}, 
it is evident that the static ES for critical ground state converges as $\sim \log^{-1} l$,
while for the quantum quench case the ES at late times demonstrates a typical $\sim l^{-1}$ dependence.
Physically, this difference can be understood from the CFT results shown in Eq.~\ref{eq:ES_W}, Eq.~\ref{eq:Wstatic} and Eq.~\ref{eq:Wdynamic}.
It is worth noting, this indicates that the ES in quantum quench process converges much faster than that of critical ground state. 
(For example, in dynamics $l=10$ (the smallest size we consider) gives $1/l = 0.1$ , 
while in static case $l=256$ (the largest size we consider) gives $1/ \log l = 0.18$.)
Such slow convergence of the ES for critical ground state is difficult to give reliable conformal information (see below). 
By comparison, the ES in the quantum quench dynamics is easy to reach convergence for extracting the conformal tower structures.

\begin{table}
	\caption{\label{tab:op}
		A comparison of the operator content in the Ising CFT ($c=1/2$) and the scaling of the ES for static (critical ground state) and dynamic (quantum quench) case. 
		Here $\Delta$ and $D$ are conformal dimension and degeneracy respectively.}
	\begin{ruledtabular}
		\begin{tabular}{ccp{0.8cm}<{\centering}p{0.8cm}<{\centering}p{1cm}<{\centering}p{1cm}<{\centering}p{1cm}<{\centering}p{1cm}<{\centering}}
			$i$-th & \multirow{2}*{sector} & \multicolumn{2}{c}{CFT} & \multicolumn{2}{c}{dynamic ES} & \multicolumn{2}{c}{static ES} \\
			level & & $\Delta$ & $D$ & $\Delta$ & $D$ & $\Delta$ & $D$ \\
			\hline
			\\[-1.5 ex]
			3 & $\epsilon$ & $3/2$ & 1 & 1.500    & 1 & 1.430 & 1 \\
			4 & $I$        & 2     & 1 & 2.000    & 1 & 1.930 & 1 \\
			5 & $\epsilon$ & $5/2$ & 1 & 2.501    & 1 & 2.300 & 1 \\
			6 & $I$        & 3     & 1 & 3.001    & 1 & 2.800 & 1 \\
			7 & $\epsilon$ & $7/2$ & 1 & 3.503    & 1 & 3.135 & 1 \\
			8 & $I$        & 4     & 2 & 4.001(3) & 2 & 3.635 & 1 \\
			9 & $\epsilon$ & $9/2$ & 2 & 4.501(5) & 2 & 3.729 & 1 \\
		\end{tabular}
	\end{ruledtabular}
\end{table}

Second, the ES in dynamic process perfectly converges to 
the CFT expectation, however, the ES of critical ground state does not.
In Table~\ref{tab:op}, we list the operator content in Ising CFT and the numerical results.
In particular, within the numerical uncertainty, the ES of dynamic equilibrium state matches the tower structure of the CFT,
for both the conformal dimension $\Delta_i$ and its degeneracy.
As a comparison, under the proper scaling, the ES of critical ground state is fail to give conformal information.  
This can be attributed to the following reasons: 
1) The scaling of the ES converges very slowly as $\log^{-1} l$, 
which hinders a clear scaled results in the limit of $l \rightarrow \infty$;
2) The ES of static critical ground state is strongly influenced by the envelop function in entanglement Hamiltonian (see Eq. \ref{eq:GSHE}).

\begin{figure}
	\centering
	\includegraphics[width=0.8\columnwidth]{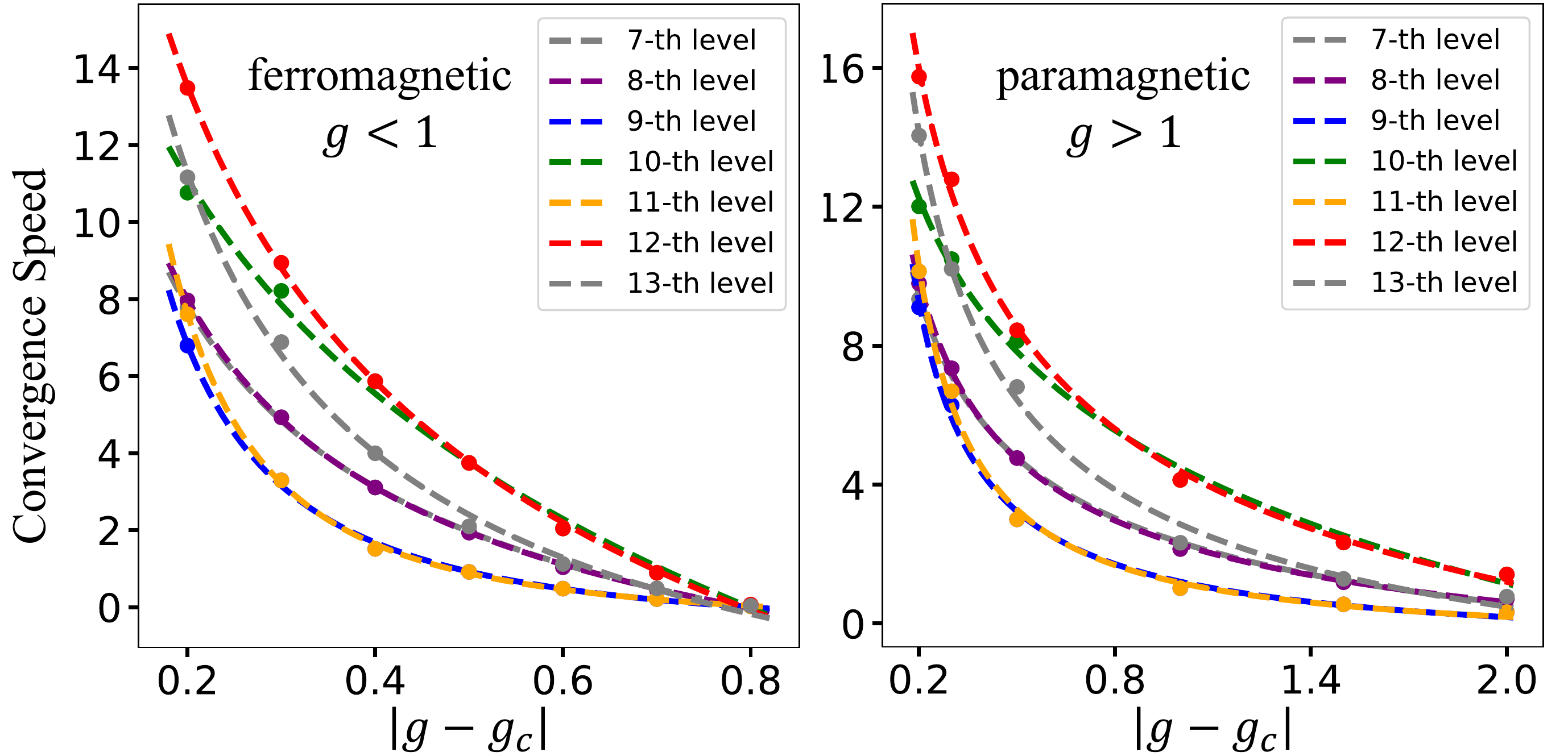}
	\caption{\label{fig:g_dependence} Dependence of convergence speed to CFT scaling operators on the initial $g$: 
	for ferromagnetic (left panel) and paramagnetic (right panel) cases.
	Dots represent the result of convergence speed extracted from the finite-size scaling, and the dash lines are result of fitting $v_{\rm converge} = a|g-g_c|^{\alpha}+b$.
    Here 7- to 13-th levels are plotted. }
\end{figure}

Third, we stress that the scaling form of the ES is robust, 
independent of the details of quench dynamics (see Ref. \cite{sm}).
Interestingly, it is found that the convergence speed (slope of the finite-size scaling function of ES)
indeed relies on the initial conditions. Here we define the convergence speed
$v_{\rm converge} = \frac{E_i-E_j}{\Delta_i-\Delta_j}/F(l)$, where $F(l)$ is the scaling function.
In the case of global quench, we have $F(l) = l^{-1}$ and $E_i-E_j = \frac{2\pi^2 (\Delta_i-\Delta_j)}{W} = \frac{\beta \pi (\Delta_i-\Delta_j)}{l}$,
which gives $v_{\rm converge} = \beta \pi$.
To check the above argument, $v_{\rm converge}$ is extracted  
and plotted in Fig.~\ref{fig:g_dependence} with the fitting in form of $\sim |g-g_c|^{\alpha}$.
The results of quenching from ferromagnetic ($g < 1$) and paramagnetic ($g > 1$) 
phases show very similar behaviors, and partially support that $v_{\rm converge}$  
is determined by effective temperature. Especially,
the differences between different levels become negligible when approaching infinite temperature limit.

\textit{Summary and discussion.---}
We have investigated the evolution of entanglement spectra during a global quench.
As suggested by boundary conformal field theory, we conclusively show that,
the entanglement spectra approaches the thermodynamic limit following $\propto l^{-1}$, where $l$ is typical subsystem size.
As a comparison, the convergence of entanglement spectra of critical ground state follows $ \propto \log^{-1} l$, much slower than that of dynamic case.
In particular, the entanglement spectra of dynamic equilibrium state encodes the conformal dimensions, 
which ambiguously pins down the nature of quantum criticality.

These results indicate that, at least in principle, one could obtain critical entanglement content in quantum dynamics, based on finite-size calculations.
As an example, we apply the TEBD method on the TFI model, and compare the results with exact solutions in Fig. \ref{fig:scaling_ES}.
This additional test implies that, if the correct scaling form is properly applied,
numerical solvers on the limited system sizes are potentially able to resolve CFT information. 
In addition, as we discussed in Fig. \ref{fig:g_dependence} and Ref.\cite{sm}, a fast convergence is possible under optimized quenching parameters.
In this context, the out-of-equilibrium scaling form  
paves a promising road for future study using various numerical methods (e.g. time-dependent denstiy-matrix renormalization group).

This work opens a number of open questions that are deserved to study in future. For example, 
it would be important to promote our findings to more systems, such as other CFT minimal models and non-integrable models, and related work is still in progress. 
How the scaling behaviors changes under the influence of the emergent gauge field and fractionalization 
will be another interesting topic.

\textit{Note Added:} In the final stage of this work, 
we became aware of a related work \href{https://arxiv.org/abs/1909.07381}{arXiv:1909.07381} (Ref. \cite{Tonni2019})
discussing the dynamics of entanglement spectra.

W.Z. thank Xueda Wen for fruitful discussion and 
Y. C. He for collaboration on a related project.  
This work is supported by start-up funding from Westlake University
and by National Natural Science Foundation of China under project 11974288.

\bibliography{dynamical_ising_CFT}

\clearpage
\widetext
\appendix \label{appendix}
\section{Conformal mapping of quantum quenching a semi-infinite line}
In order to introduce conformal mapping of global quantum quenching a semi-infinite line, first we need to represent the time evolution of density matrix $\rho(t)$ geometrically.
The initial state $|\psi_0\rangle$ is chosen to be a short-range correlated state with correlation length $\beta$ much less than the total system size, which can be considered as the ground state of a gapped Hamiltonian.
In such a choice of initial state, the system is expected to be thermalized to a finite temperature $T = 1/\beta$ at long time limit.
A much clearer description is to assume that $|\psi_0\rangle$ can be written in the form $|\psi_0\rangle \sim e^{-\frac{\beta}{4} H_{\rm CFT}}|b\rangle$, where $|b\rangle$ is a conformal boundary state satisfying $(T(x)-\overline{T}(x))|b\rangle=0$~\cite{Cardy_2016review, Calabrese_2016}.
The factor $e^{-\frac{\beta}{4} H_{\rm CFT}}$ can be considered as a conformal mapping to the boundary state $|b\rangle$, giving the free energy $F = \frac{\pi c l}{6\beta^2}$.
Technically, this is the origin of the appearing effective temperature in quantum dynamics, with $\rho(t) \sim e^{-iHt}e^{-\frac{\beta}{4} H_{\rm CFT}}|b\rangle \langle b| e^{-\frac{\beta}{4} H_{\rm CFT}} e^{+iHt}$.
It is worth noting that this assumption of thermalization is very strong, and is not always true.
The first insight is that the dynamical conserved free energy $F$ has the same form with the finite-temperature thermalization~\cite{blote1986conformal, affleck1986universal}.
Moreover, the reduced density matrix is found to be exponentially close to a thermal Gibbs state, once the interval falls inside the horizon~\cite{Cardy_2016review}.
This fact strongly support the assumption we made.
Based on above argument, we conclude that the global quench of a semi-infinite line can be described by as an infinite half-strip as shown in the left panel of Fig.~\ref{fig:geometry}.

There, in fact, are more problems about the CFT calculation.
The partial trace, which is required to calculate the reduced density matrix, will result the branch cut along $C=\left\{z=x+i\tau; x \in [0,l], \tau \in [-\frac{\beta}{4}, \frac{\beta}{4}] \right\}$.
A small disc around the entangling points $z_0=\left\{l+i\tau; \tau \in [-\frac{\beta}{4}, \frac{\beta}{4}] \right\}$ will lead to the \emph{ultraviolet} divergence, and need be removed for regularization.
In BCFT, a normal way is to introduce a boundary state $|a\rangle$, imposing around the entangling points.
The regulator $|a\rangle$ will raise a boundary term known as the Affleck-Ludwig boundary entropy~\cite{affleck1991universal}, which is also interesting to investigate.
After this operation, the branch cut $C$ becomes a surface connecting $|a\rangle$ and $|b\rangle$.
It is important to note that the geometry of an infinite half-strip, including the branch cut caused by the partial trace, is topologically equivalent to an annulus (cylinder without boundaries).
One can build a conformal mapping from the original infinite half-strip to an annulus, where the two boundary states $|a\rangle$ and $|b\rangle$ locate at two edges, connecting by the mapped branch cut.
In such a geometry, the entanglement Hamiltonian can be considered as the generator of translation, so it could be a good choice.

The conformal mapping $w=f(z)$, from the infinite half-strip in $z$-plane to an annulus in $w$-plane, can be achieved through following there steps.
First, by $\xi(z) = \sinh (\frac{2\pi z}{\beta})$, the infinite half-strip in $z$-plane is mapped to the right half part of $\xi$ plane.
Note that the entangling points $z_0$ are mapped to $\xi_0=\xi(z_0)$.
Second, we map the entangling points $\xi_0$ to ${\rm Re}(\xi'_0) = (0, +\infty)$ by $\xi \to \xi'(\xi)=\frac{1+\overline{\xi_0}}{1+\xi_0} \cdot \frac{\xi+\overline{\xi_0}}{\xi-\xi_0}$.
Third, applying $w(\xi')=\log(\xi')$, the right half $\xi'$ plane is mapped to an annulus, with circumference $2\pi$ along $v={\rm Im}w$ direction and width $W$ along the $u={\rm Re}w$ direction.
The entangling points are simply removed, and the branch cut $C$ is mapped to a curve $f(C)$ connecting the two edges of the annulus (the two boundary states $|a\rangle$ and $|b\rangle$).

\section{Calculation of the entanglement Hamiltonian}
Once we build up the conformal mapping $w=f(z)$ to an annulus, the entanglement Hamiltonian on original geometry can be written as a local integral over the Hamiltonian density $H(x)$.
Remember that, on the annulus in the $w$-plane, the entanglement Hamiltonian $H_E$ can be considered as the generator of translation along the $v={\rm Im}w$ direction, as
\begin{equation}
H_E = -2\pi \int_{v={\rm const}} T_{vv} du = 2\pi \left [ \int_{f(C)} T(w)dw + \int_{\overline{f(C)}} \overline{T}(\overline{w})d\overline{w} \right ] \ ,
\end{equation}
where $f(C)$ is the branch cut $C$ after conformal mapping, and the Hamiltonian density $T_{00}=-T_{vv}$ is written in terms of the holomorphic and anti-holomorphic components $T(w)$ and $\overline{T}(\overline{w})$.
After mapping back to the $z$-plane, we have
\begin{equation}\label{eq:HECFT}
H_E = 2\pi \left [ \int_C \frac{T(z)}{f'(z)} dz + \int_{\overline{C}} \frac{\overline{T}(\overline{z})}{\overline{f'(z)}} dz \right ] \ .
\end{equation}
Using equation~\ref{eq:HECFT}, it is straightforward to obtain the entanglement Hamiltonian $H_E(t)$ in our considered case (after analytical continuation $\tau \to it$), as
\begin{equation}\label{eq:TDHE}
\begin{aligned}
H_E(t) \ = \ & 2\beta \int_0^l \frac{\sinh \left[ \frac{\pi(x-l)}{\beta} \right] \cosh \left[ \frac{\pi(x-2t+l)}{\beta} \right] \sinh \left[ \frac{\pi(x+l)}{\beta} \right] \cosh \left[ \frac{\pi(x-2t-l)}{\beta} \right] }{\cosh \left( \frac{2\pi t}{\beta} \right) \sinh \left( \frac{2\pi l}{\beta} \right) \cosh \left[ \frac{2\pi(x-t)}{\beta} \right]} T(x,t) dx \\
& + 2\beta \int_0^l \frac{\sinh \left[ \frac{\pi(x-l)}{\beta} \right] \cosh \left[ \frac{\pi(x+2t+l)}{\beta} \right] \sinh \left[ \frac{\pi(x+l)}{\beta} \right] \cosh \left[ \frac{\pi(x+2t-l)}{\beta} \right] }{\cosh \left( \frac{2\pi t}{\beta} \right) \sinh \left( \frac{2\pi l}{\beta} \right) \cosh \left[ \frac{2\pi(x+t)}{\beta} \right]} \overline{T}(x,t) dx \ .
\end{aligned}
\end{equation}

Simply taking $t=0$ in equation~\ref{eq:TDHE}, one can obtain the entanglement Hamiltonian in equilibrium
\begin{equation}
\begin{aligned}
H_E(t=0) \ = \ & 2\beta \int_0^l \frac{\sinh \left[ \frac{\pi (x-l)}{\beta} \right] \cosh \left[ \frac{\pi (x-l)}{\beta} \right] \sinh \left[ \frac{\pi (x+l)}{\beta} \right] \cosh \left[ \frac{\pi (x+l)}{\beta} \right]}{\sinh \left( \frac{2\pi l}{\beta} \right) \cosh \left( \frac{2\pi x}{\beta} \right)} T_{00}(x) dx \\
= \ & \beta \int_0^l \frac{\sinh \left[ \frac{\pi (x-l)}{\beta} \right] \sinh \left[ \frac{\pi (x+l)}{\beta} \right]}{\sinh \left[ \frac{2\pi (l+x)}{\beta} \right] + \sinh \left[ \frac{2\pi (l-x)}{\beta} \right]} T_{00}(x) dx \\
= \ & \beta \int_0^l \left\{ \sinh^{-1} \left[ \frac{2\pi (x-l)}{\beta} \right] - \sinh^{-1} \left[ \frac{2\pi (x+l)}{\beta} \right] \right\}^{-1} T_{00}(x) dx \ .
\end{aligned}
\end{equation}
An important limit is to take $\beta \to \infty$, i.e. the critical ground state.
In this case, the above equation becomes
\begin{equation}
H_E (t=0)= \int_A \frac{\pi(l^2-x^2)}{l} H(x)dx \ ,
\end{equation}
which implies that the entanglement Hamiltonian on lattice geometry has a different structure with the CFT Hamiltonian.

For quenching to long time $t \to \infty$, the entanglement Hamiltonian
\begin{equation}
\begin{aligned}
H_E(t\to \infty) \ = \ 2\beta \int_0^l \frac{\sinh \left[\frac{\pi(l-x)}{\beta}\right] \sinh \left[\frac{\pi(l+x)}{\beta}\right]}{\sinh \frac{2\pi l}{\beta}} T_{00}(x) dx \simeq \beta \int_0^l T_{00}(x) dx,
\end{aligned}
\end{equation}
shares the same structure with CFT Hamiltonian up to a global factor $\beta$.

\section{Scaling behavior of the entanglement spectrum and entropy: from the width $W$ of the mapped annulus}
In this section we show that the width $W$ along the $u={\rm Re}w$ direction of the mapped annulus plays important role in entanglement spectrum and entropy.
The width $W$ can be expressed as $W = {\rm Re}(\mathcal{W}) = \frac{1}{2}(\mathcal{W} + \overline{\mathcal{W}})$ with $\mathcal{W}=f(i\tau+l-\epsilon) -f(i\tau)$, where $f(z)$ is the conformal mapping from the original infinite half-strip to the annulus.
A straightforward calculation (after analytical continuation to real time $\tau \to it$) gives
\begin{equation}
W = \log \left\{
\frac{2 \sinh [\frac{2\pi (l-\epsilon/2)}{\beta}] \cosh (\frac{2\pi t}{\beta})}
{\sinh (\frac{2\pi \epsilon / 2}{\beta}) \sqrt{2\cosh(\frac{4\pi l}{\beta}) + 2\cosh (\frac{4\pi t}{\beta})}}
\right\} 
\end{equation}
By taking the long-time lime $t \to \infty$, one can obtain its thermal value
\begin{equation}
W_{\rm thermal} \coloneqq W(t \to \infty) \ = \log \frac{\sinh [\frac{2\pi (l-\epsilon/2)}{\beta}]}{\sinh (\frac{2\pi \epsilon/2}{\beta})} \simeq \frac{2\pi}{\beta} l
\end{equation}
The entanglement spectrum has the structure
\begin{equation}
E_i - E_j = \frac{2\pi^2(\Delta_i - \Delta_j)}{W_{\rm thermal}} \simeq \frac{\beta \pi}{l} (\Delta_i - \Delta_j)
\end{equation}
where $E_i$ is the $i$-th level of entanglement Hamiltonian, and $\Delta_i$ is the level of $i$-th scaling operator. This also gives the long-time entropy as
\begin{equation}
S(t \to \infty) \simeq \frac{\pi c}{3 \beta} l
\end{equation}
For the time after reaching the saturate time $t = l$, $W$ can be calculated by expanding to the term in $t$, straightforward algebra results
\begin{equation}
W(t>L) \simeq W_{\rm thermal} - \frac{1}{2} \exp [-\frac{4\pi(t-l)}{\beta}]
\end{equation}
which shows an exponential approaching to thermalization.

\section{Lattice effect: inhomogeneous effective temperature $\beta$}

The lattice effect plays important role in realization of dynamical CFT in lattice models.
In this section, we show the effective temperature $\beta$ is not uniform in lattice models, and strongly influences the behavior of dynamic entanglement spectrum.
Consider the long time limit of the global quench, which is described by a conformal mapping to the annulus (cylinder without boundaries).
The same geometry can also describe the finite-temperature thermalization.
Recall that the thermal density matrix with inverse temperature $\beta$ has the form
\begin{equation}
\rho_{\rm thermal} = \frac{1}{Z} e^{-\beta H} = \frac{1}{Z} e^{-\sum_{k}\beta \epsilon_k \eta_k^\dagger \eta_k} \ ,
\end{equation}
where the (integrable) Hamiltonian can be diagonalized in the momentum space as $H=\sum_{k}\epsilon_k \eta_k^\dagger \eta_k$, and $Z$ is the normalization factor.
In our case, the reduced density matrix can be written in a similar form
\begin{equation}
\rho_A = \frac{1}{\tilde{Z}} e^{-\sum_{k}\varepsilon_k \xi_k^\dagger \xi_k} \ .
\end{equation}
A direct comparison results a mode dependent effective temperature
\begin{equation}
\beta_k = \varepsilon_k / \epsilon_k \ .
\end{equation}
The entanglement spectra $\{ E_i \}$, i.e. the eigenvalues of $-\log \left[ \rho_A \right]$, are simply
\begin{equation}
E_i = -\log \left[ \frac{1}{\tilde{Z}} e^{-\sum_{k}\varepsilon_k n_k^i} \right]
= \sum_{k}\varepsilon_k n_k^i + \log \left[ {\tilde{Z}} \right]
= \sum_{k} \beta_k \epsilon_k n_k^i + \log \left[ {\tilde{Z}} \right]  \ ,
\end{equation}
where the occupation numbers $n_k^i = 0, 1$.
it is worth noting that, the renormalization factor ${\tilde{Z}}$, also the infinite order entropy $S_A^{(n \to \infty)} = E_1 = \log \left[ {\tilde{Z}} \right]$, is effectively coupled to all modes in the momentum space, as ${\tilde{Z}} = \prod_k (1+e^{-\varepsilon_k})$.
However, the Schmidt gap $E_2 - E_1$ is always dependent only on one $\beta_k$ corresponding to the lowest level in $\left\{ \varepsilon_k \right\}$.
The above argument explains why the effective temperature $\beta$ is inhomogeneous at different levels of entanglement spectra.
Moreover, through a derivation for Gaussian model, Calabrese and Cardy~\cite{Calabrese_2007} show that $\beta$ becomes independent on $k$ when the correlation length (inverse mass) of the initial state $\to 0$.
In our case, the mass term does not appear in the Hamiltonian directly, but 
the correlation length $\beta$ decreases with increasing the distance between $g$ and $g_c=1$.
Therefore an initial state with $g$ far away from the critical point $g_c=1$ is expected to give a better result in numeric, we will show that this is the case in following section.

\section{Dependence on the initial $g$ of the global quench}

\begin{figure}
	\centering
	\includegraphics[width=0.8\columnwidth]{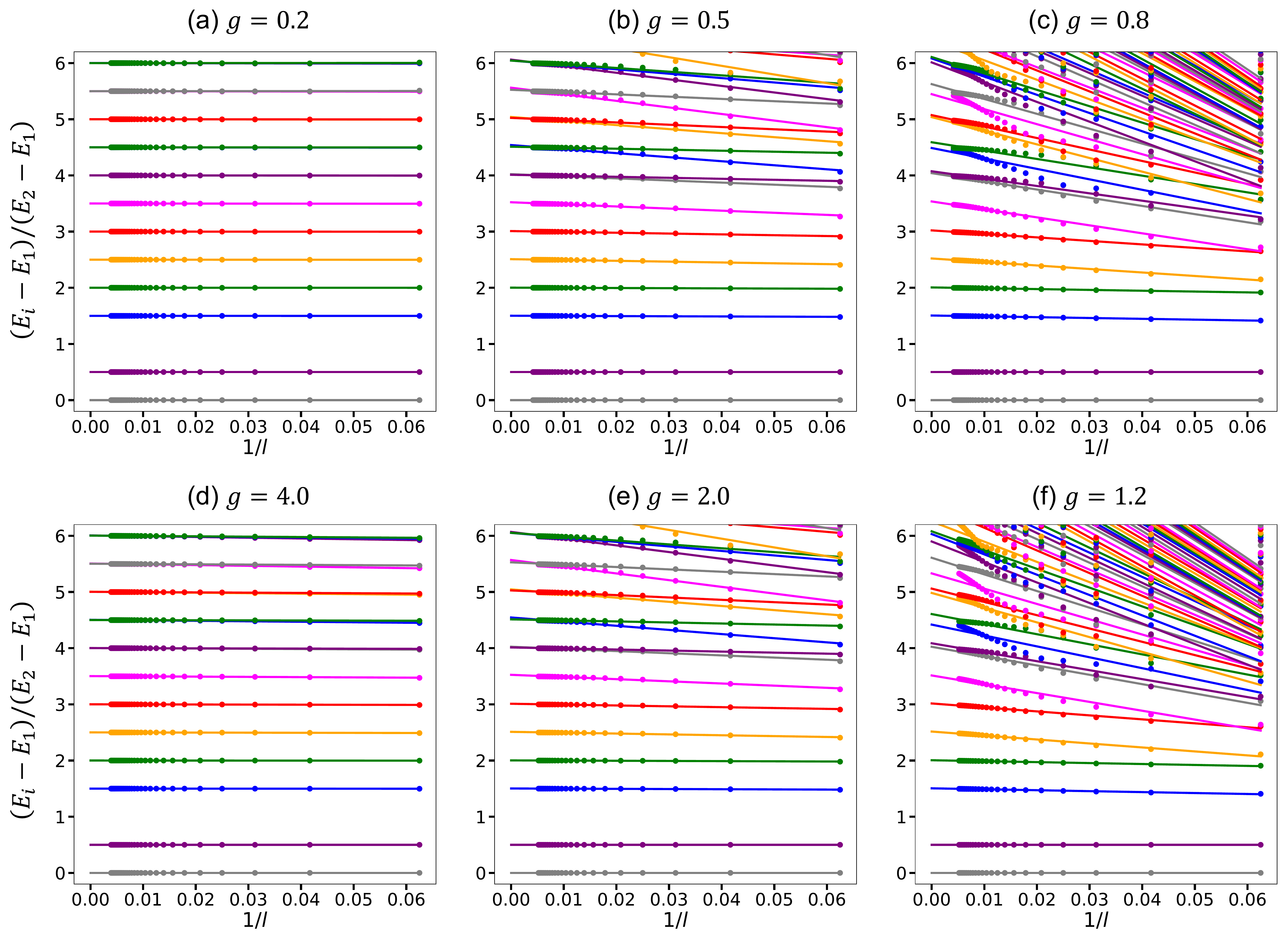}
	\caption{\label{fig:g_fit_tower} Finite-size scaling of entanglement spectrum of dynamic equilibrium state for different initial $g$, with scaling function $F(l)=l^{-1}$. Both cases of ferromagnetic $g<$ (top) and paramagnetic $g>1$ (bottom) are presented.}
\end{figure}

In this section we present numeric of dynamic entanglement spectrum for different initial $g$.
A finite-size scaling of the numerical results is shown in Fig.~\ref{fig:g_fit_tower}, and the same behaviors are observed in $g<1$ ferromagnetic and $g>1$ paramagnetic cases.
As proposed in the last section and the main text, the convergence of entanglement spectrum exhibits a dependence on the initial condition (the distance between initial $g$ and the critical point $g_c=1$) of the global quench.
The case of initial $g=4$, presented in the main text, is a typical example of short-range correlated state with $\beta \ll L$.
A very quick convergence in finite-size scaling can be directly observed.
As shown in Fig.~\ref{fig:g_fit_tower}(a) and (d), when the initial $g$ is far away from the critical point, the slope of scaling function $F(l) = l^{-1}$ shares the same value for different levels.
When the initial $g$ becomes closer to the critical point, especially when $g=0.8$ and $1.2$, the scaling function $F(l) = l^{-1}$ even not works for dynamic entanglement spectrum, since the initial state is no longer a short-range correlated state.
In Table.~\ref{tab:g_op}, we list the numerical results after finite-size scaling for different initial $g$.
As we argued, when initial $g$ is closer to the critical point, i.e. the gap is smaller, the numerical results are inconsistent with the CFT prediction.

\begin{table}[b]
	\caption{\label{tab:g_op}
		A comparison of the operator content in dynamic equilibrium state during global quench from different initial $g$.
		Here $\Delta$ and $D$ are the eigenvalue and degeneracy of $i$-th level respectively.}
	\begin{ruledtabular}
		\begin{tabular}{ccp{0.8cm}<{\centering}p{0.8cm}<{\centering}p{1cm}<{\centering}p{1cm}<{\centering}p{1cm}<{\centering}p{1cm}<{\centering}p{1cm}<{\centering}p{1cm}<{\centering}p{1cm}<{\centering}p{1cm}<{\centering}p{1cm}<{\centering}p{1cm}<{\centering}p{1cm}<{\centering}p{1cm}<{\centering}}
			$i$-th & \multirow{2}*{sector} & \multicolumn{2}{c}{CFT} & \multicolumn{2}{c}{$g=0.2$} & \multicolumn{2}{c}{$g=0.5$} & \multicolumn{2}{c}{$g=0.8$} & \multicolumn{2}{c}{$g=1.2$} & \multicolumn{2}{c}{$g=2.0$} & \multicolumn{2}{c}{$g=4.0$} \\
			level & & $\Delta$ & $D$ & $\Delta$ & $D$ & $\Delta$ & $D$ & $\Delta$ & $D$ & $\Delta$ & $D$ & $\Delta$ & $D$ & $\Delta$ & $D$\\
			\hline
			\\[-1.5 ex]
			3  & $\epsilon$ & $3/2$  & 1  & 1.50       & 1 & 1.50    & 1 & 1.51       & 1 & 1.51       & 1 & 1.50     & 1 & 1.50    & 1 \\
			4  & $I$ & 2 & 1              & 2.00       & 1 & 2.00    & 1 & 2.01       & 1 & 2.01       & 1 & 2.00     & 1 & 2.00    & 1 \\
			5  & $\epsilon$ & $5/2$  & 1  & 2.50       & 1 & 2.51    & 1 & 2.52       & 1 & 2.52       & 1 & 2.51     & 1 & 2.50    & 1 \\
			6  & $I$ & 3 & 1              & 3.00       & 1 & 3.01    & 1 & 3.02       & 1 & 3.02       & 1 & 3.01     & 1 & 3.00    & 1 \\
			7  & $\epsilon$ & $7/2$  & 1  & 3.50       & 1 & 3.52    & 1 & 3.54       & 1 & 3.52       & 1 & 3.53     & 1 & 3.50    & 1 \\
			8  & $I$ & 4 & 2              & 4.00       & 2 & 4.01(2) & 2 & 4.04(7)    & 2 & 4.03(9)    & 2 & 4.01(3)  & 2 & 4.00    & 2 \\
			9  & $\epsilon$ & $9/2$  & 2  & 4.50       & 2 & 4.51(4) & 2 & 4.49(59)   & 2 & 4.42(61)   & 2 & 4.51(5)  & 2 & 4.50    & 2 \\
			10 & $I$ & 5 & 2              & 5.00       & 2 & 5.02(4) & 2 & 5.04(7)    & 2 & 4.98(5.06) & 2 & 5.03(5)  & 2 & 5.00    & 2 \\
			11 & $\epsilon$ & $11/2$ & 2  & 5.50       & 2 & 5.52(6) & 2 & 5.45(63)   & 2 & 5.33(61)   & 2 & 5.53(7)  & 2 & 5.50(1) & 2 \\
			12 & $I$ & 6 & 3              & 6.00       & 3 & 6.04(6) & 3 & 6.01(11)   & 3 & 5.90(6.08) & 3 & 6.05(7)  & 3 & 6.00(1) & 3 \\
			13 & $\epsilon$ & $13/2$ & 3  & 6.46(50)   & 3 & 6.54(7) & 3 & 6.40       & 1 & 6.26       & 1 & 6.55(7)  & 3 & 6.50(1) & 3 \\
			14 & $I$ & 7 & 3              & 7.00       & 3 & 7.07    & 3 & 6.62(5)    & 2 & 6.55(60)   & 1 & 7.07(9)  & 3 & 7.01    & 3 \\
			15 & $\epsilon$ & $15/2$ & 4  & 7.49(50)   & 4 & 7.57    & 4 & 6.93(7.09) & 3 & 6.78       & 1 & 7.57(9)  & 4 & 7.50(1) & 4 \\
			16 & $I$ & 8 & 5              & 7.99(8.00) & 5 & 8.06(9) & 5 & 7.33       & 1 & 6.93(9)    & 2 & 8.06(10) & 5 & 8.00(1) & 5 \\
		\end{tabular}
	\end{ruledtabular}
\end{table}

\end{document}